\documentclass[12]{report}
\usepackage{graphicx}
\usepackage{cancel}
\usepackage{amsmath}    
\long\def\symbolfootnote[#1]#2{\begingroup\def\thefootnote{\fnsymbol{footnote}}
\footnote[#1]{#2}\endgroup}

\begin{document}
\setcounter{page}{0}
\title{Elementary Considerations on the Interpretation of the Foundations of Quantum Mechanics\footnote{This paper, whose original title was ``Elementare \"Uberlegungen zur Interpretation der Grundlagen der Quanten-Mechanik",  has been translated from the German by Dileep Karanth, Department of Physics, University of Wisconsin-Parkside, Kenosha, USA.}}
\author{Albert Einstein, \\ Institute for Advanced Study, Princeton, N.J.}
\date{\today}
\maketitle

\begin{abstract}
This paper appeared in a collection of papers titled ``Scientific Papers Presented to Max Born on his
retirement from the Tait Chair of Natural Philosophy  in the University of Edinburgh'', published in 1953 (Oliver
and Boyd), pages 33-40.
\end{abstract}

What I find odd about the present situation is as follows: There are no doubts about the mathematical formalism of the theory, but about the physical interpretation of its statements. What is the relationship of the  $\psi$-function to the facts of a concrete single case, that is, to the individual situation of an isolated system? In other words: What does the $\psi$-function say about the (individual) ``real-state"? 

Then one can wonder, whether one can assign any meaning to this question. One can of course assume the following standpoint: Only the result of an individual observation is ``Real", not a result objectively existing in time and space, independently of the act of observation. When one takes this clear positivist standpoint, one does not need to give any thought to the question what the ``real state" is to be taken as, within the framework of the quantum theory. The effort then seems like a fencing-match with a ghost. 

This neatly positivist standpoint, when logically carried forward, has a irreparable weakness: it leads to the conclusion that all verbally expressible statements are declared to be devoid of meaning. Does one have a right to declare that the result of an individual observation makes sense, that is, is true or false? Can such a description not be based on falsehood, or experiences that we can regard as recollections of a dream or as hallucinations? Does the difference between waking and dream experiences have any objective sense at all? 
Finally, what remain as ``real" are only the experiences of a residual ``I", without any possibility of saying anything at all about them, since they turn out to be senseless without exception, in the concepts used in statements, according to the positivist approach. 

In reality, the independent concepts and conceptual systems used in our statements are human creations, self-made tools, whose validation and value ultimately rests on the fact that they help us to order our experiences ``with convenience" (\emph{turn out to be true}\symbolfootnote[1]{Translator's Footnote: Emphasis added by translator. The original text simply has the word  Bew\"ahrung in parentheses}). Expressed otherwise, these tools are justified to the extent that make it possible to ``explain"\footnote{At the basis of the linguistic affinity of the concepts ``true" and ``turn out to be true" lies a fundamental relationship. This idea must not be misunderstood in a utilitarian sense.\\ Translator's Footnote: The original text simply has the words ``wahr" and ``sich bew\"ahren" in quotes.} experiences. 

The validity of concepts and conceptual systems should be judged from the standpoint of ``turning out to be true" alone. This applies also to the concepts ``physical reality", ``reality of the external world", ``real state of a system". A priori there is no justification to postulate them as being logically necessary, or to rule them out. It is only whether they turn out to be true, that decides their validity. Behind these word-symbols stands a programme, which has gained unquestioned standing with the development of physical thinking until the establishment of the quantum theory. It should all be reduced to conceptual objects from the space-time sphere, and to the legitimate relationships that are supposed to be valid for them. Nothing enters this description which applies to an empirical knowledge regarding these objects. To the moon at all times a position in space (with respect to a coordinate system in use) is attributed, independent of whether perceptions of this position exist or not. This is the kind of description one has in mind, when one speaks of the physical description of the ``external world", whatever the choice of the elementary building blocks (material points, field, etc.) that may be taken as the basis of such a description. 

The validation of this programme was not seriously doubted by physicists so long as it seemed that everything entering into such a description, could  in principle be confirmed empirically in every case. That this was an illusion was first proven in the realm of quantum phenomena by Heisenberg in a manner that was convincing to physicists. 

The concept ``physical reality" was now felt to be problematic, and the question arose as to what it was exactly that theoretical physics (through quantum mechanics) sought to describe, and to what the laws that physics established applied to. This question was answered variously. 

To arrive nearer to an answer, we consider what quantum mechanics has to say about macro-systems, that is, about such objects, which we experience as ``directly perceivable". About such objects we know that they and the laws that apply to them can be represented by classical physics to great, if not unlimited, accuracy. We do not doubt that for such objects at all times, there is a real space configuration (position) as well as a velocity (a momentum), that is, a \textit{real situation} -- all with the approximation required by the quantum structure. 

We ask: Does quantum mechanics (with the expected approximations) imply the real description provided by classical mechanics for macroscopic objects? Or if this question cannot be answered with a simple ``Yes", in which sense is this the case? We will reflect on these questions with a specific example.

\section*{The Specific Example}
The system, consisting of a bullet of around 1 mm in diameter, moving to and fro between two parallel walls, which are about 1 meter apart (along the x-axis of a certain coordinate system.) The collisions are supposed to be elastic. In this idealized macro-system, we regard the walls as being replaced by ``sharply" decaying potential energy expressions, into which enter only the coordinates of the material particles constituting the bullet. Care is taken to ensure that the processes of reflection do not give rise to any coupling between the center-of-mass coordinate $x$ of the bullet and ``inner coordinates" (including the angle coordinates). Thus we are able to specify the position of the bullet (aside from its radius) through $x$ alone, for our purposes. 

In the quantum-mechanical sense, a process of well-defined energy is involved. The de Broglie wave ($\psi$-function) is harmonic in the time coordinate. Further, it differs from $0$ only between $ x = - \frac{l}{2} $ and $ x = \frac{l}{2} $. At the endpoints of the path, continuity with the vanishing of the $\psi$-function outside the path is achieved by the requirement that for $ x = \pm \frac{l}{2}$, it must be true that $\psi = 0$. 

The $\psi$-function is then a standing wave, which inside the path can be represented as the superposition of two harmonic waves propagating in opposite directions: 
\begin{subequations}
\begin{equation}
\psi = \frac{1}{2} A \exp^{ i(at -bx) } + \frac{1}{2} A \exp^{ i(at + bx) }
\end{equation}
or 
\begin{equation}
\psi =  A \exp^{iat} \cos bx
\end{equation}
\end{subequations}
From (1a) we see that the factor A in both terms must be chosen to be the same, so that the boundary conditions at the ends can be satisfied. A can be chosen to be real without loss of generality. b is determined according to the Schr\"odinger equation,  via the mass m. We regard the factor A as normalized in the well-known manner.

In order that a comparison with the example with the corresponding classical problem be fruitful, we must also require that the de Broglie wavelength $\frac{2 \pi}{b}$ be small in comparison to $l$. 

First of all, as is customary in the light of Born's probabilistic interpretation of the significance of the $\psi$-function, we set:
\begin{center}
\begin{equation*}
W = \int \psi \bar{\psi} dx = A^2 \int \cos^2 (bx)dx
\end{equation*}
\end{center}
That is the probability that the center-of-mass coordinate $x$ of the bullet lies in a given interval $\Delta x$. Apart from an undulatory ``fine-structure", whose physical reality is certain, it is a simple constant times $\Delta x$.

What about the probabilities of the values of momentum and velocities  of the bullet? These probabilities are obtained from the Fourier-decomposition of $\psi$. If equation (1) were valid from $ - \infty$ to $ \infty $, then it would already have been the desired Fourier-decomposition. There would then have been two equally probable, sharply defined momenta which would have been equal but opposite. However, since both wavetrains are bounded, the term they each contribute is a continuum Fourier-decomposition whose spectral region is narrower, the greater the value of the de Broglie wavelength contained in the region $l$. This follows from the fact that only two nearly sharp equal and opposite values of momentum are possible - which values incidentally coincide with that of the classical case, and which both have equal probability. 

Both these statistical results are apart from the small variations which are required by the quantum structure, which apply to a time ensemble of systems in the case of a classical theory. In this respect the theory is fully satisfactory. 

However we now ask ourselves: Can this theory provide a real description of an individual case? This question must be answered with a ``No". In deciding this it is important that we are dealing with a ``Macro-system". Since for a macro-system we are sure that it is at any time in a ``real state" which is described approximately correctly by classical mechanics. The individual macro-system of the kind we have considered also has a nearly well-defined center-of-mass coordinate at all times -- at least averaged over a small time interval -- and a nearly well-defined momentum (determined with respect to the choice). Neither of these two pieces of information can be arrived at via the $\psi$-function  (1). From the $\psi$-function, only such information can be extracted which pertain to a \textit{statistical ensemble} of systems of the type under consideration. 

That for a macro-system under consideration, not every $\psi$-function satisfying the Schr\"odinger equation approximately corresponds to the real description in a classical mechanical sense, becomes particularly obvious when we consider a $\psi$-function, which is made up of the superposition of two solutions of type (1), but whose frequencies (or energies) differ considerably from one another. Then such a superposition does not correspond to any real classical mechanical state at all (but to a statistical ensemble of such real states in the sense of Born's interpretation.)

Summing it all up, we conclude: Quantum Mechanics describes ensembles of systems, not an individual system. In this sense, the description by means of the $\psi$-function is an incomplete description of a single system, and not a description of its real state. 

Remark: The following objection can be raised against this conclusion: The case we have considered of extreme sharpness of frequency is a limiting case, for which probably the requirement of similarity with a classical mechanical problem can break down. When one allows a finite if small region of frequency in time, then by means of a suitable choice of amplitudes and phases of $\psi$-functions, one can arrive at a $\psi$-function that is approximately localized in position and momentum. Can one not try to limit the allowed $\psi$-functions according to this viewpoint, and arrange for the allowed $\psi$-functions to be a representation of a single system?

Such a possibility must immediately be rejected on the grounds that the position-localization of such a representation cannot be achieved for all times. 

The fact that the Schr\"odinger equation combined with the Born interpretation does not lead to a description of the real state of a single system, naturally gives rise to a search for a theory which is free of this limitation. 

So far there have been two attempts in this direction, which share the features that they maintain the Schr\"odinger equation, and give up the Born interpretation.  The first effort goes back to de Broglie and has been pursued further by Bohm with great perspicacity. 

As Schr\"odinger's original investigation of the wave equation by analogy (linearizing of Jacobi's Equation of Analytical Mechanics) leads to classical mechanics, so by analogy should the equation of motion of a quantized single system -- supported by a solution $\psi$ of the Schr\"odinger equation -- be established. The rule is as follows: $\psi$ should be cast in the form

\begin{center}
\begin{equation*}
\psi = R e^{iS}
\end{equation*}
\end{center}

Thus out of $\psi$ are derived the (real) functions of the coordinates $R$ and $S$. Differentiation of $S$ with respect to the coordinates gives the momenta or velocities of the system as functions of time, if the coordinates are known for a definite value of time in a given coordinate system.

A look at (1a) shows that in our case $ \frac{\partial S}{\partial x} $ vanishes, and with it the velocity. This objection, which incidentally was raised already a quarter-century ago by Pauli against this theoretical attempt, in very serious when brought to bear on our example. The vanishing of the velocity contradicts the well-founded requirement that in the case of a macroscopic system, the motion must approximately agree with the classical mechanical motion.

The second attempt, which aims at achieving a real description of an individual system, based on the Schr\"odinger equation, has been made by Schr\"odinger himself. Briefly, his ideas are as follows. The $\psi$-function itself represents reality, and does not stand in need of the Born interpretation. The atomic structures about which the $\psi$-function is supposed to carry information, do not exist at all, at least, not as localized structures.  Carried over to our macroscopic system, it means that: macroscopic objects do not exist as such. In any case, there is no such thing -- even approximately -- as the position of a center-of-mass at a definite time. Here also the demand that the quantum theoretical description of the motion of a macroscopic system be approximately in agreement with the corresponding classical mechanical description, is not respected. 

The result of our considerations is this: The acceptable interpretation of the Schr\"odinger equation is the statistical interpretation given by Born. However this does not provide any real description of an individual system, but only statistical predictions about ensembles of systems. 

In my opinion it is in principle not satisfactory that such a theoretical framework should lie at the foundations of physics, especially since the objective describability of an individual macroscopic system (description of the ``real state") cannot be given up without the picture of the physical world  dissolving in a cloud, as it were. Finally, the judgement that physics must strive towards a real description of an individual system, is unavoidable. Nature as a whole can only thought of as an individual (singly existing) system and not as an ``ensemble of systems". 

\section*{Translator's Acknowledgments}
The translator is grateful to the Estate of Albert Einstein at the Hebrew University of Jerusalem, and to Princeton University Press for permission to translate this paper, and to publish the translation on the web. 
This project was first suggested to the translator by his supervisor Dr. Daniel Kennefick, Assistant Professor, Dept. of Physics, University of Arkansas, Fayetteville. The translator is also grateful to Dr. Kerry Magruder, Curator, and to Dr. JoAnn Palmeri, Librarian, History of Science Collections, University of Oklahoma Libraries for making the original article available to him, during his stay in Norman, facilitated by an award of the Andrew Mellon Travel Fellowship. 

\section*{Translator's Note}
This translation (or any other translation) may not be reproduced in print, except with the permission of the Estate of Albert Einstein at the Hebrew University of Jerusalem, and of Princeton University Press (for English-language rights). 
The translator requests readers to inform him of any mistakes in the translation, and also to make suggestions  for improving the translation.  
\end{document}